\documentclass{ws-book975x65edit} 
\usepackage{ws-book-har}

\newcommand{\beq}{\begin{equation}}
\newcommand{\eeq}{\end{equation}}
\newcommand{\beqa}{\begin{eqnarray}}
\newcommand{\eeqa}{\end{eqnarray}}
\newcommand{\om}{\Omega_m}

\newcommand{\gscr}{{\mathcal{G}}}
\newcommand{\vscr}{{\mathcal{V}}}

\begin{document} 

\bibliographystyle{ws-book-har} 

\chapter{Frontiers of Dark Energy} 
\author{Eric V.\ Linder \\
Berkeley Lab \& University of California, Berkeley, CA 94720, USA \\ 
Institute for the Early Universe, Ewha Womans University, Seoul, Korea} 
\date{\today}

\section{Introduction to Dark Energy} \label{sec:intro}

Emptiness -- the vacuum -- is a surprisingly rich concept in cosmology. 
A universe devoid of all matter and radiation can still have evolution 
in space and time.  In fact there can be very distinct empty universes, 
defined through their geometry.  One of the great modern quests in science 
is to understand the hidden constituents of the universe, neither matter 
nor radiation, and their intimate relation with the nature of the quantum 
vacuum and the structure of spacetime itself. 

Cosmologists are just beginning to probe the properties of the cosmic 
vacuum and its role in reversing the attractive pull of gravity to cause 
an acceleration in the expansion of the cosmos.  The cause of this 
acceleration is given the generic name of dark energy, whether it is 
due to a true vacuum, a false, temporary vacuum, or a new relation between 
the vacuum and the force of gravity.  Despite the common name, the 
distinction between these origins is of utmost interest and physicists 
are actively engaged in finding ways to use cosmological observations 
to distinguish which is the true, new physics.  See \citet{caldbook} 
and \citet{hutbook} in this volume for further details on the theoretical 
origins and observational probes, respectively, and 
\citet{calkam,durmaar,friehuttur,silvestri} for other recent reviews. 

Here we will discuss how to relate the theoretical ideas to the experimental 
constraints, how to understand the influences of dark energy on the 
expansion and structure in the universe, and what frontiers of new physics 
are being illuminated by current and near-term data.  In Sec.~\ref{sec:dyn} 
we consider the vacuum, quantum fields, and their interaction with material 
components.  The current level of our understanding about the properties 
of dark energy is reviewed in Sec.~\ref{sec:curr}, and we relate this to 
a few, robust theories for the origin of dark energy.  Looking to the 
frontiers of exploration, Sec.~\ref{sec:front} anticipates what we may 
learn from experiments just now underway or being developed.

\section{The Dynamics of Nothing \label{sec:dyn}} 

Emptiness, in general relativity, merely means that nothing has been 
put on the stage of space and time.  The framework, however, the 
structure of space and time and their relation into spacetime, is 
part of the theory itself.  A universe devoid of matter, radiation, 
all material contents still has geometry.  We will consider here only 
the highly symmetrical case of a simply connected universe (no holes 
or handles) that is homogeneous (uniform among spatial volumes) and 
isotropic (uniform among spatial directions).  A universe with only 
spatial curvature is called a Milne universe, or often just an empty 
universe.  If even spatial curvature vanishes, then this is a Minkowski 
universe, a relativistic generalization of Euclidean space.  

Suppose we now consider an energy completely uniform everywhere 
in space.  One possibility for this is the energy of the spatial curvature 
itself, for example in the Milne universe.  In evolving toward a lower 
energy state, the universe reduces the curvature energy, proportional to 
the inverse square radius of curvature, $a^{-2}$, by expanding.  That is, 
the factor $a$ increases with time (and is often called the expansion 
factor or scale factor).  Since the dynamical timescale of a 
self-gravitating system is proportional to the inverse square root of 
the energy density, then $a\propto t$.  We see that there is no 
acceleration, i.e.\ $\ddot a=0$, and the expansion continues at the 
same rate, $\dot a=$ constant forever. 

Now imagine another uniform energy not associated with spatial 
curvature: a vacuum energy, a nonzero ground state level of 
energy.  If this were negative, it would reduce the curvature energy and 
could counteract the expansion, possibly even causing collapse of the 
universe.  That is, $a$ reduces with time until it reaches zero. 
Such a uniform energy is called a negative cosmological constant.  
However suppose the vacuum energy were positive: then it would add 
to the energy and the expansion, increasing the rate such that we 
have $\ddot a>0$ -- acceleration. 

Finally, remove the spatial curvature completely.  With just the 
positive cosmological constant one still has the acceleration (recall 
the spatial curvature did not contribute to the acceleration positively 
or negatively).  Nothing is in the universe but a positive, uniform 
energy.  This is called a de Sitter universe.  Most interesting, though, 
is what happens when we restore the matter and radiation into the picture. 
Matter and radiation have the usual gravitational attraction that pulls 
objects together, fighting against expansion.  They act to decelerate 
the expansion.  Depending on the relative contributions then between 
matter etc.\ and the vacuum, the final result can be either a decelerating 
or accelerating universe.  One of the great paradigm shifts in cosmology 
was the realization and experimental discovery \citep{perl99,riess98} that 
we live in a universe that accelerates in its expansion, where gravity is 
not predominantly attractive. 

This is really quite striking a development, opening up whole frontiers 
of new physics.  At its most personal, it reminds us of the ``principle 
of cosmic modesty''.  Julius Caesar (at least through George Bernard 
Shaw) defined a barbarian as one who 
``thinks that the customs of his tribe and island are the laws of nature.'' 
After Copernicus we have moved beyond thinking the Earth is the center 
of the universe; with the development of astronomy we know that the 
Milky Way Galaxy is not the center of the universe; through physical 
cosmology we know that what we are made of -- baryons and leptons -- 
is not typical of the matter in the universe; and now we even realize 
that the gravitational attraction we take as commonplace is not the 
dominant behavior in the universe.  We are decidedly on the doorstep 
of new physics. 

How then do we elucidate the role of the vacuum?  A first step is 
certainly to determine whether we are indeed dealing with a uniform, 
constant energy filling space.  The vacuum is the lowest energy state 
of a quantum field.  One can picture this as a field of harmonic 
oscillators, imaginary springs at every point in space, and ask whether 
these springs are identical and frozen, or whether they have some 
spatial variation and motion.  An assemblage of values defined at 
points in space and time is basically a scalar field, and we seek to 
know whether dark energy is a true cosmological constant or a dynamical 
entity, perhaps one whose energy is not in the true ground state but is 
temporarily lifted above zero and is changing with time. 

The scalar field approach is a fruitful one since one can use it as an 
effective description of the background dynamics of the universe 
even if the origin of acceleration is from another 
cause.  That is, one can define an effective energy density and 
effective pressure (determining how the energy density changes with time), 
and use that in the equations governing the expansion (although the 
growth of inhomogeneities can be influenced by other degrees of freedom). 
This description of the cosmic expansion holds even if there is no 
physical field at all, such as in the case of a modification of the 
gravitational theory.  (We discuss some of the ways to distinguish 
between explanations in \S\ref{sec:front}.) 

Indeed, it is instructive to review some historical cases where dynamics 
indicated new physics beyond what was then known.  In the 18th century, 
the motion of the planet Uranus did not accord with predictions of 
Newton's laws of gravitation applied from the other material contents of the 
solar system.  Two choices presented themselves: the laws were inadequate, 
or the knowledge of the material contents was incomplete.  Keeping the 
laws intact and asking what new material content was needed to explain 
the anomaly led to the discovery of Neptune.  In the 19th century, the 
motion of Mercury disagreed with the laws and material contents known. 
While some again sought a new planet, Einstein developed extensions to 
Newtonian gravity -- the solution lay in new laws.  For dark energy, we 
do not know whether we need to add new contents -- a quantum scalar field, 
say -- or an extension to Einstein gravity.  However what is certain is 
that we are in the midst of a revolution in physics.  While Einstein's 
correction to Mercury's orbit led to a minuscule $43''$/century of extra 
precession, dark energy turns cosmology upside down by changing 
gravitational attraction into accelerated expansion, dominates the 
expansion rate, and determines the ultimate fate of the universe. 

We can investigate dark energy's dynamical influence in more mathematical 
detail through the scalar field language (without assuming a true, 
physical scalar field).  The Lagrangian density for a scalar field is just 
\beq 
{\mathcal L}=\frac{1}{2}\phi_{;\mu}\phi^{;\mu}+V(\phi)\,, 
\eeq 
where $\phi$ is the value of the field, $V$ is its potential, and 
$;\mu$ denotes derivatives with respect to the time and space coordinates. 
Using the Noether construction of the energy-momentum tensor, one can 
identify the energy density $\rho$ and isotropic pressure $p$ (all other 
terms vanishing under homogeneity and isotropy) as 
\beqa 
\rho&=&\frac{1}{2}\dot\phi^2+\frac{1}{2}(\nabla\phi)^2+V\\ 
p&=&\frac{1}{2}\dot\phi^2+\zeta\,(\nabla\phi)^2-V\,. 
\eeqa 
Here $\zeta=-1/6$ (1/2) depending on whether the field is treated 
as spatially incoherent or coherent.  If the spatial gradient 
terms dominated, then the pressure to density ratio would be $-1/3$ 
(i.e.\ acting like spatial curvature) or $+1$ (i.e.\ acting like a 
stiff fluid or gradient tilt) in the two cases.  However, in the vast 
majority of cases the spatial gradients are small compared to the 
other terms and are neglected. 

It is convenient to discuss the scalar field properties in terms of 
the equation of state parameter 
\beq 
w\equiv \frac{p}{\rho} = \frac{(1/2)\dot\phi^2-V}{(1/2)\dot\phi^2+V}\,, 
\eeq 
where the first equality is general and we neglect spatial gradients 
in the second equality, as in the rest of the article.  When the kinetic 
energy term dominates, then $w$ approaches $+1$; when the potential energy 
dominates, then $w\to-1$, and when they balance (as in oscillating around 
the minimum of a quadratic potential) then $w=0$, like nonrelativistic 
matter.  Acceleration occurs when the total equation of state, the 
weighted sum (by energy density) of the equations of state of each 
component, is $w_{\rm tot}<-1/3$.  The Friedmann equation for the 
acceleration of the expansion factor is 
\beq 
\frac{\ddot a}{a}=-4\pi G\,(\rho_{\rm tot}+3p_{\rm tot})=-4\pi G \sum 
\rho_w\,(1+3w)\,, 
\eeq 
where $G$ is Newton's constant and we set the speed of light equal to 
unity. 

The other equation of motion is either the Friedmann expansion equation 
\beq 
H^2\equiv \frac{\dot a^2}{a^2}=\frac{8\pi G}{3}\rho_{\rm tot}\,, 
\eeq 
where we can include any curvature energy density in $\rho_{\rm tot}$, 
or the energy conservation or continuity equation 
\beq 
\dot\rho=-3H(\rho+p) \qquad {\rm or}\qquad \frac{d\ln\rho}{d\ln a}=-3(1+w)\,. 
\eeq 
The continuity equation holds separately for each individually conserved 
component.  In particular, for a scalar field we can write the continuity 
equation as a Klein-Gordon equation 
\beq 
\ddot\phi+3H\dot\phi+dV/d\phi=0\,. 
\eeq 

To examine the dynamics of the dark energy, one can solve for $w(a)$; 
it is also often instructive to work in the phase space of $w'$-$w$, 
where a prime denotes $d/d\ln a$.  For example, many models can be 
categorized as either thawers or freezers \citep{caldlin}: their behavior 
either starts with the field frozen by the Hubble friction of the 
expanding universe (so kinetic energy is negligible and $w=-1$), and 
then at late times the field begins to roll, moving $w$ away from $-1$, 
or the field starts off rolling and gradually comes to settle at a minimum 
of the potential, asymptotically reaching $w=-1$. 

Since the dark energy does not always dominate the energy budget and 
expansion of the universe, it is also useful to examine the dynamics of 
the full system of components.  One can define variables representing 
each contribution to the energy density, say, and obtain a coupled system 
of equations \citep{copelid}.  For example, for a scalar field 
\beqa 
x'&=&-3x+\lambda\sqrt{\frac{3}{2}}y^2+\frac{3}{2}x\,[2x^2+(1+w_b)(1-x^2-y^2)] 
\label{eq:xdyn}\\ 
y'&=&-\lambda\sqrt{\frac{3}{2}}xy+\frac{3}{2}y\,[2x^2+(1+w_b)(1-x^2-y^2)] 
\label{eq:ydyn} \,, 
\eeqa 
where $x=\sqrt{\kappa\dot\phi^2/(2H^2)}$, $y=\sqrt{\kappa V/H^2}$, 
$\kappa=8\pi G/3$, 
and $\lambda=-(1/V)dV/d(\phi\sqrt{3\kappa})$, with $w_b$ being the equation 
of state of the background, dominating component (e.g.\ matter, with $w_b=0$, 
during the matter dominated era).  To solve these equations one must specify 
initial conditions and the form of $V(\phi)$, i.e.\ $\lambda$. 

\begin{figure}[!htb]
  \begin{center}
\psfig{file=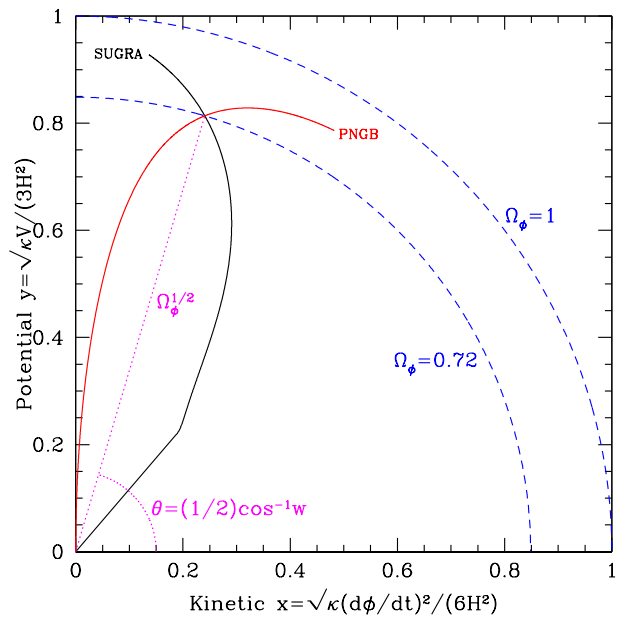,width=8.5cm} 
\caption{The dynamics in the $y$-$x$, or potential-kinetic energy phase 
space for a thawing (PNGB) and a freezing (SUGRA) field.  The dark energy 
density $\Omega_\phi$ acts as a radial coordinate, while the dark energy 
equation of state $w$ acts as an angular coordinate.  Note the constancy 
of $w$ (i.e.\ the angle $\theta$) for SUGRA at early times, 
when it is on the attractor trajectory.  The models have 
been chosen to have the same values today, $\Omega_{\phi,0}=0.72$ and 
$w_0=-0.839$ (where the curves cross).  The curves end in the future at 
$a=1.47$.} 
  \label{fig:oww}
  \end{center}
\end{figure}

The fractional dark energy density $\Omega_w=x^2+y^2$, so its evolution 
is bounded within the first quadrant of the unit circle in the $x$-$y$ 
plane (taking $\dot\phi>0$; it is simple enough to generalize the equations), 
and the dark energy equation of state is $w=(x^2-y^2)/(x^2+y^2)$.  
So the dynamics can be represented in polar coordinates, with the density 
being the radial coordinate and the equation of state the angular coordinate 
(twice the angle with respect to the $x$ axis is $2\theta=\cos^{-1}w$). 
Figure~\ref{fig:oww} illustrates some dynamics in the $y$-$x$ energy 
density component (or $w$-$\Omega_\phi$) plane.

\begin{figure}[!htb]
  \begin{center}
\psfig{file=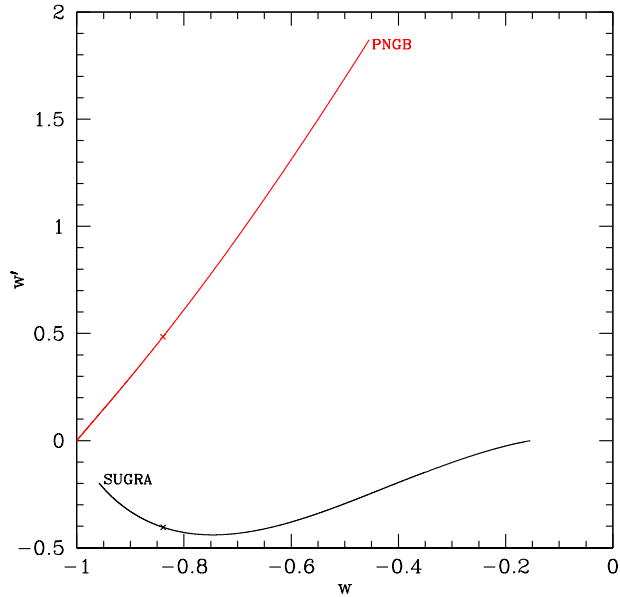,width=8.5cm} 
  \caption{The dynamics in the $w'$-$w$ phase plane for a thawing 
(PNGB) and a freezing (SUGRA) field. 
The right or left curvature in Fig.~\ref{fig:oww} here translates into 
$w'>0$ or $<0$.  
The thawer starts in a frozen 
state ($w=-1$, $w'=0$) and evolves away from the cosmological constant 
behavior, while the freezer starts at some constant $w$ given by 
an attractor solution and then evolves as its energy density becomes 
more substantial, eventually approaching the cosmological constant 
state.  The x's mark the present state, and the curves end in the 
future at $a=1.47$.} 
  \label{fig:wwp}
  \end{center}
\end{figure}

The term in square brackets in Eqs.~(\ref{eq:xdyn}) and (\ref{eq:ydyn}) 
is simply $1+w_{\rm tot}$.  Another way of viewing the dynamics is through 
the variation of the dark energy equation of state 
\beq 
w'=-3(1-w^2)+\lambda(1-w)x\sqrt{2}\,. 
\eeq 
One can readily see that $w=-1$ (and hence $x=0$) is a fixed point, with 
$w'=0$.  It can either be a stable attractor (in the case of freezing 
fields) or unstable (in the case of thawing fields).  Figure~\ref{fig:wwp} 
illustrates some dynamics in the phase plane $w'$-$w$, an alternate 
view to Figure~\ref{fig:oww}.  Considerably more 
detail about classes of dynamics is given in \citet{caldlin,paths}. 
For example, through nonstandard kinetic terms one can get dynamics with 
$w<-1$, sometimes called phantom fields \citep{bigrip}.

The dynamical view of dark energy identifies several key properties 
that would lead to insight into the nature of the physics behind 
acceleration.  Since $w=-1$ is a special state, we can ask whether 
the dark energy always stays there, i.e.\ is it a cosmological constant? 
Does dark energy act like a thawing (roughly $w'>0$) or freezing 
(roughly $w'<0$) field?  Is it ever phantom ($w<-1$)? 
One could also look back at the continuity equation and ask whether 
each component is separately conserved or whether there is interaction. 
Keeping overall energy conservation, one could write 
\beqa 
\frac{d\ln\rho_w}{d\ln a}&=&-3(1+w)+\frac{\Gamma}{H}\\ 
\frac{d\ln\rho_m}{d\ln a}&=&-3(1+w_m)-\frac{\Gamma}{H}\,, 
\eeqa 
for the dark energy and (dark) matter components, where $\Gamma$ 
represents the interaction.  The impact of this is to shift each 
equation of state, such that $w_{\rm eff}=w-\Gamma/(3H)$ and 
$w_{m,{\rm eff}}=w_m+\Gamma/(3H)$.

Such interactions act as a fifth force violating the Equivalence 
Principle if dark energy responds to different components in 
different ways.  Certainly interaction with baryons is highly 
constrained otherwise we would have found dark energy from particle 
physics experiments.  The shift in equation of state could make 
dark energy that intrinsically has $w>-1$ look like a phantom field, 
and vice versa (see \citet{wei} for some current constraints).  
Dynamical analysis does allow us to make some general 
statements: for example, consider a phantom field arising from a 
negative kinetic term.  The dynamical variable $y=\sqrt{\kappa V/H^2}$ 
has a fixed point when $y'_c=0$, so the potential obeys 
$V'/V=2H'/H\equiv 3(1+w_{\rm tot})$.  However, such negative kinetic term 
fields roll {\it up\/} the potential so $V'$ is positive.  Therefore 
$w_{\rm tot}$ must be less than $-1$ and the field must remain 
asymptotically phantom, even in the presence of interactions.

\section{Knowing Nothing \label{sec:curr}} 

The existence of dark energy was first discovered through the geometric 
probe of the distance-redshift relation of Type Ia supernovae 
\citep{perl99,riess98}.  Such data have been greatly expanded and refined 
so that now the analysis of the Union2 compilation of supernova data 
\citep{union2}, together with other probes, establishes that 
the energy density contribution of dark energy to the total energy 
density is $\Omega_{de}=0.719\pm0.017$ and the dark energy equation of 
state, or pressure to density ratio, is $w=-1.03\pm0.09$ (assumed constant).  

Other cosmological probes are now investigating cosmic acceleration, 
although none by themselves have approached the leverage of supernovae. 
Experiments underway use Type Ia and Type II supernovae, 
baryon acoustic oscillations, cosmic microwave background measurements, 
weak gravitational lensing, and galaxy clusters with the Sunyaev-Zel'dovich 
effect and X-rays.  See \citet{hutbook} for more detailed discussion. 

\begin{figure}[!htbp]
  \begin{center}
\psfig{file=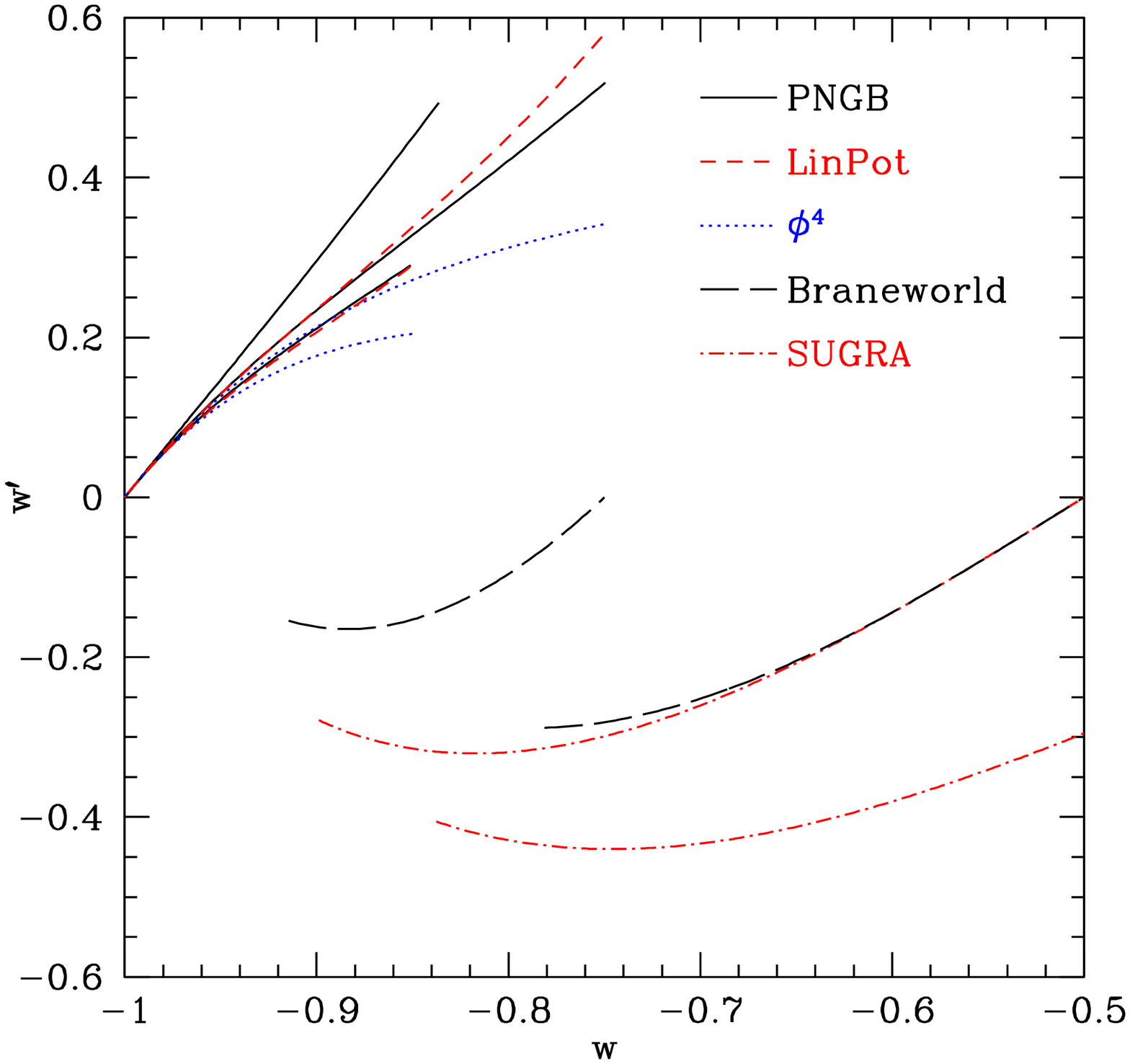,width=7.4cm}\\ 
\psfig{file=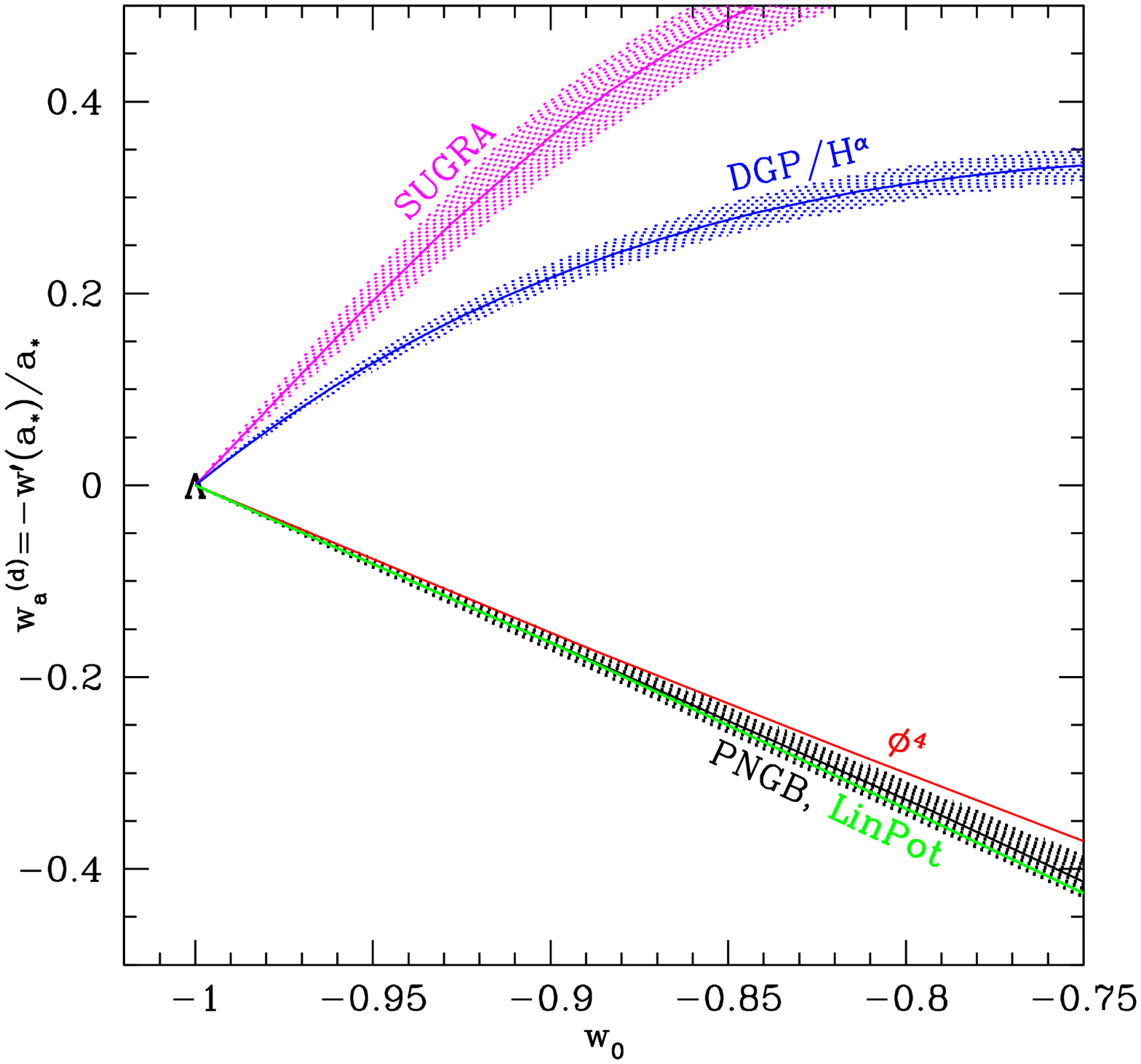,width=7.4cm}
  \caption{[Top panel] Representative models exhibiting a diversity of 
dynamics are plotted for various parameter values in the $w$-$w'$ phase 
space, including braneworld/$H^\alpha$ models ($\alpha=1$ DGP and 
$\alpha=0.5$). 
[Bottom panel] Using the calibrated dark energy parameters $w_0$ and 
$w_a$, dark energy models and families lie in tightly homogeneous regions.  
Contrast this with the top panel, showing the same models before 
calibration (note $w_a$ has the opposite sign from $w'$).  
We here vary over all parameters in the potentials.  
Shading shows the effect of scanning over $\pm0.03$ in $\om$ (we omit 
the shading for $\phi^4$ and linear potential models to minimize confusion; 
the width would be about half that shown for PNGB).  
Distinctions between thawing and freezing models, and between freezing 
models, become highlighted with calibration. From \citet{calde}.} 
  \label{fig:thawfcalib}
  \end{center}
\end{figure}

Observables such as the distance-redshift relation and Hubble 
parameter-redshift relation, and those that depend on these in a more 
complex manner, can be used to test specific models of dark energy. 
For some examples of this, see \citet{beylam,tdavissdss,mhh2}.  However it 
is frequently useful to have a more model independent method of 
constraining dark energy properties.  We have already seen in the previous 
section that one can classify many models into the general behaviors of 
thawers and freezers.  There appears diversity within each of these 
classes, but \citet{calde} found a calibration relation between the dark 
energy equation of state value and its time variation that defines 
homogeneous families of dark energy physics.  Figure~\ref{fig:thawfcalib} 
illustrates both the diversity and the calibration.

This calibration provides a physical basis for a very simple but powerful 
relation between the equation of state value and time variation in the dark 
energy dynamics phase plane.  The resulting parametrization 
\beq 
w(a)=w_0+w_a(1-a) 
\eeq 
gives a highly accurate match to the observable relations of distance 
$d(z)$ and Hubble parameter $H(z)$.  This form, emphatically not a 
Taylor expansion, achieves $10^{-3}$ accuracy 
on the observables and matches the $w_0$-$w_a$ parametrization devised to 
fit the exact solutions for scalar field dynamics \citep{linprl}. 

\begin{figure}[!htbp]
\begin{center}
\psfig{file=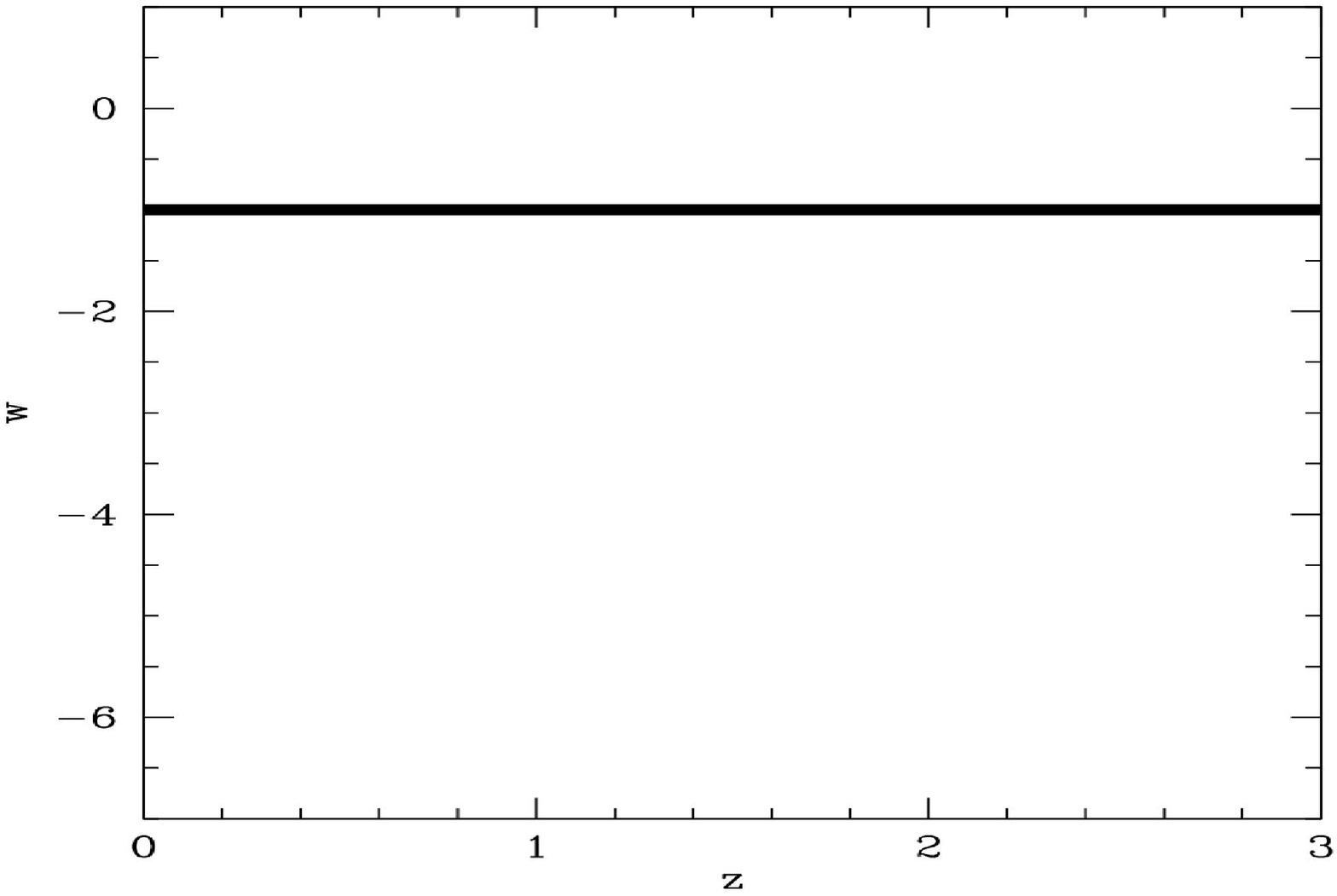,width=0.47\textwidth}
\psfig{file=wbins_3_BAO_CMB_H0wmap7.epsi,width=0.45\textwidth} \\ $\,$ \\ 
\psfig{file=wbins_3+1_BAO_CMB_H0wmap7.epsi,width=0.48\textwidth}
\psfig{file=wbins_eqerr5_BAO_CMB_H0wmap7.epsi,width=0.45\textwidth}
\caption{Constraints from the Union2 supernova compilation, 
WMAP7 CMB, SDSS DR7 baryon acoustic oscillation, and Hubble 
constant data on the dark energy equation of state $w(z)$, in 
redshift bins.  Top left plot appears to show that data 
have zeroed in on the cosmological constant value of $w=-1$, 
but this assumes $w$ is constant.  When one allows for the values 
of $w$ to be different in different redshift bins, our current 
knowledge of dark energy is seen to be far from sufficient. 
Top right plot shows that we do not yet have good constraints on 
whether $w(z)$ is constant.  Bottom left plot (note change of scale) 
shows we have little 
knowledge of dark energy behavior, or even existence, at $z>1$. 
Bottom right plot shows we have little detailed knowledge of dark energy 
behavior at $z<1$.  Outer (inner) boxes show 68\% confidence limits 
with (without) systematics.  
The results are consistent with $w=-1$, but also allow 
considerable variation in $w(z)$.  Adapted from \citet{union2}. 
}
\label{fig:wbin}
\end{center}
\end{figure}

Current data constrains $w_0$ to $\sim0.3$ and $w_a$ to $\sim1$, which 
is insufficient to answer any of the questions raised in the previous 
section, e.g.\ whether dark energy is a cosmological constant or not, 
is thawing or freezing, etc.  To give a clear picture of our current 
state of knowledge, Figure~\ref{fig:wbin} displays the constraints from 
all current data in several different ways.

For example, for $w$ held constant, \citet{union2} find that 
the energy density contribution of dark energy to the total energy 
density is $\Omega_{de}=0.719\pm0.017$ and the dark energy equation of 
state, or pressure to density ratio, is $w=-1.03\pm0.09$ (68\% confidence 
level, including systematic uncertainties). 
While viewing the constraints on $w$ under the assumption that it is 
constant (upper left panel) gives an impression of substantial precision, 
in fact none of the key physical questions have been answered.  The 
upper right panel shows that when we leave open the values of $w$ in 
different redshift ranges (redshift $z=a^{-1}-1$), then we have no 
reasonable constraints on whether $w$ is in fact constant in time.  Recall 
that for a simple scalar field, $w$ is bounded from below by $-1$, and 
must be less than $-1/3$ to provide acceleration.  So the panoply of 
current data does not give much evidence for or against constancy of $w$. 

The bottom left panel demonstrates that we have no constraints at all on 
dark energy above $z\approx1.6$, neither knowing its properties nor even 
whether it exists.  In the bottom right panel it is clear that the situation 
at low redshift (near the present time) is also quite uncertain: does $w$ 
differ from $-1$, and if so in which direction? 

On the theoretical front, no consensus exists on any clear concept for 
the origin of dark energy.  Any expansion history can be accommodated 
by a combination of potential and kinetic terms, but it is really not 
a case of an embarrassment of riches.  There are two main problems: 
any potential that one writes down should receive quantum corrections 
at high energies and so end up different from the original intent, and 
the energy scale corresponding to dark energy is much lower (by many tens 
of orders of magnitude) than scales associated with initial conditions in 
the early universe.  How do we cue dark energy and cosmic acceleration to 
appear on the stage of the universe at the right moment? 
That is, one generically requires fine tunings to describe the universe 
today starting from high energy physics. 

To surmount these difficulties requires some symmetry to preserve the 
form of the potential, and some tracking mechanism to keep dark energy 
in the wings until the proper moment.  Simple scalar fields fail on one 
or both of these counts (the cosmological constant fails on both). 
However there are a few possibilities that might offer guidance toward 
a more robust theory. 

\begin{figure}[!htbp]
\begin{center}
\psfig{file=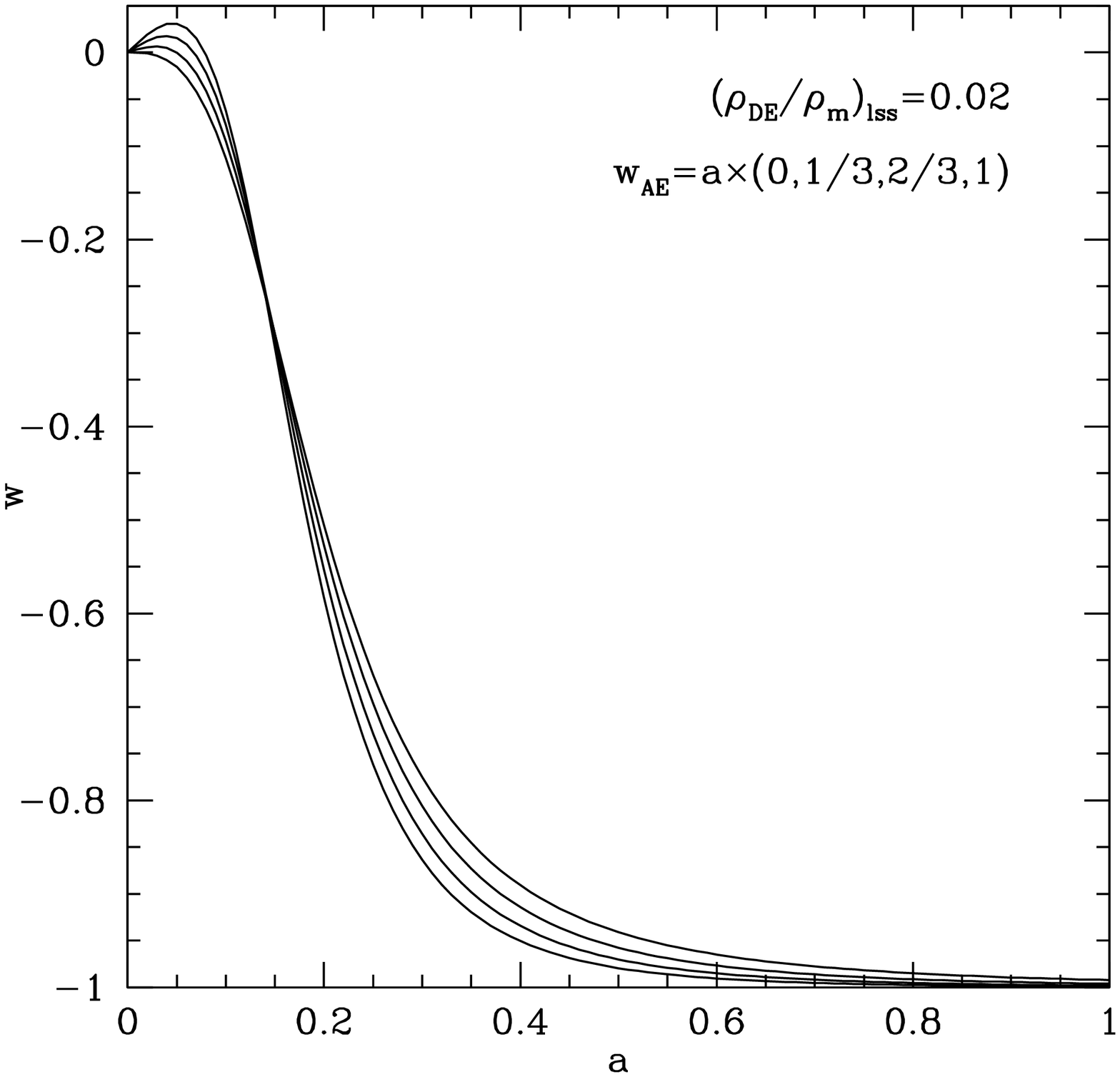,width=7.4cm}\\
\psfig{file=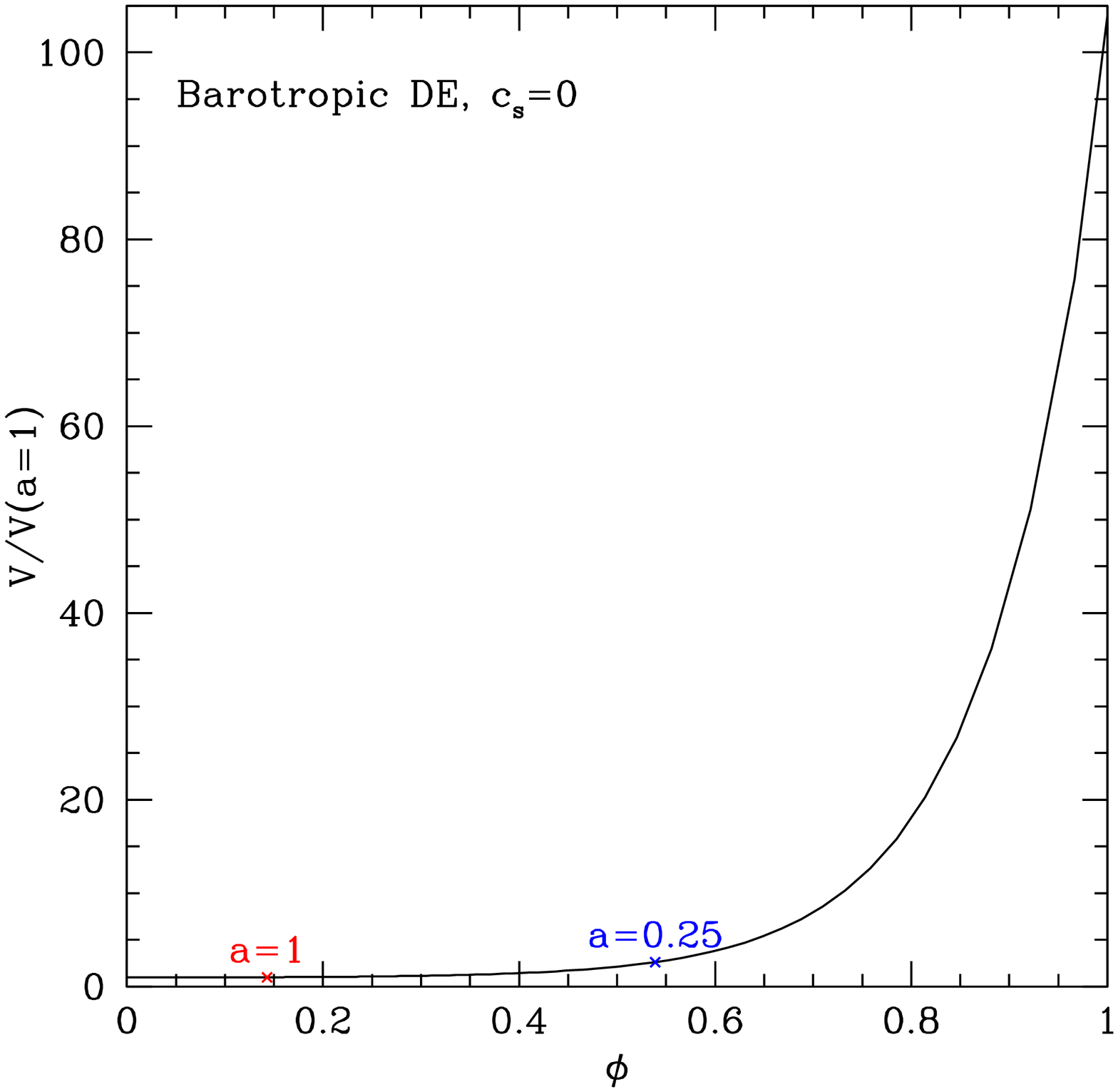,width=7.4cm}
\caption{Barotropic models make a rapid transition from $w=0$ at 
high redshift ($a\ll1$) to $w\approx-1$ more recently: the transition 
from $w=-0.1$ to $w=-0.9$ always takes less than 1.5 e-folds.  This is 
inherent in the barotropic physics and, in distinction 
to quintessence, gives a prediction that observations of the recent 
universe should find $w\approx-1$.  [Bottom panel] Effective potential 
corresponding to a barotropic model with $c_s=0$.  The x's mark where 
the field is today and at $a=0.25$, showing that it has reached the 
flat part of the potential, and so $w\approx-1$ for the last $\sim$90\% of 
the age of the universe. 
}
\label{fig:baro}
\end{center}
\end{figure}

Some theories, such as the pseudo-Nambu Goldstone boson (PNGB) model 
\citep{frie95}, impose a symmetry that protects the form.  Such 
theories are known as natural theories.  However to achieve acceleration 
at the right time still requires a restricted range of initial conditions.  
Attractor models where dark energy is kept off stage, but not too far off, 
for the radiation and matter dominated eras, are a useful class 
\citep{ratrap,wett88,zws,lidsch}.  An intriguing class of models that 
incorporates 
both these advantages is the Dirac-Born-Infeld (DBI) action based on higher 
dimension theories \citep{alisha,martindbi,ahn2,ahn1}.  This employs a 
geometric constraint to preserve the potential and a relativistic 
generalization of the usual scalar field dynamics to provide the attractor 
property.  The attraction to $w=-1$ actually occurs in the future, but 
prevents the dynamics from diverging too far from $w=-1$ at any time.  
Another class of interest, although not arising directly from high energy 
physics, is that of barotropic models.  In the barotropic aether model 
the equation of state 
naturally transitions rapidly from acting like another matter component 
to being attracted to $w=-1$, thus ``predicting'' $w=-1$ for much of the 
observational redshift range and ameliorating the coincidence of recent 
acceleration \citep{linsch}; see Figure~\ref{fig:baro}.

\begin{figure}[!htb]
  \begin{center} 
\psfig{file=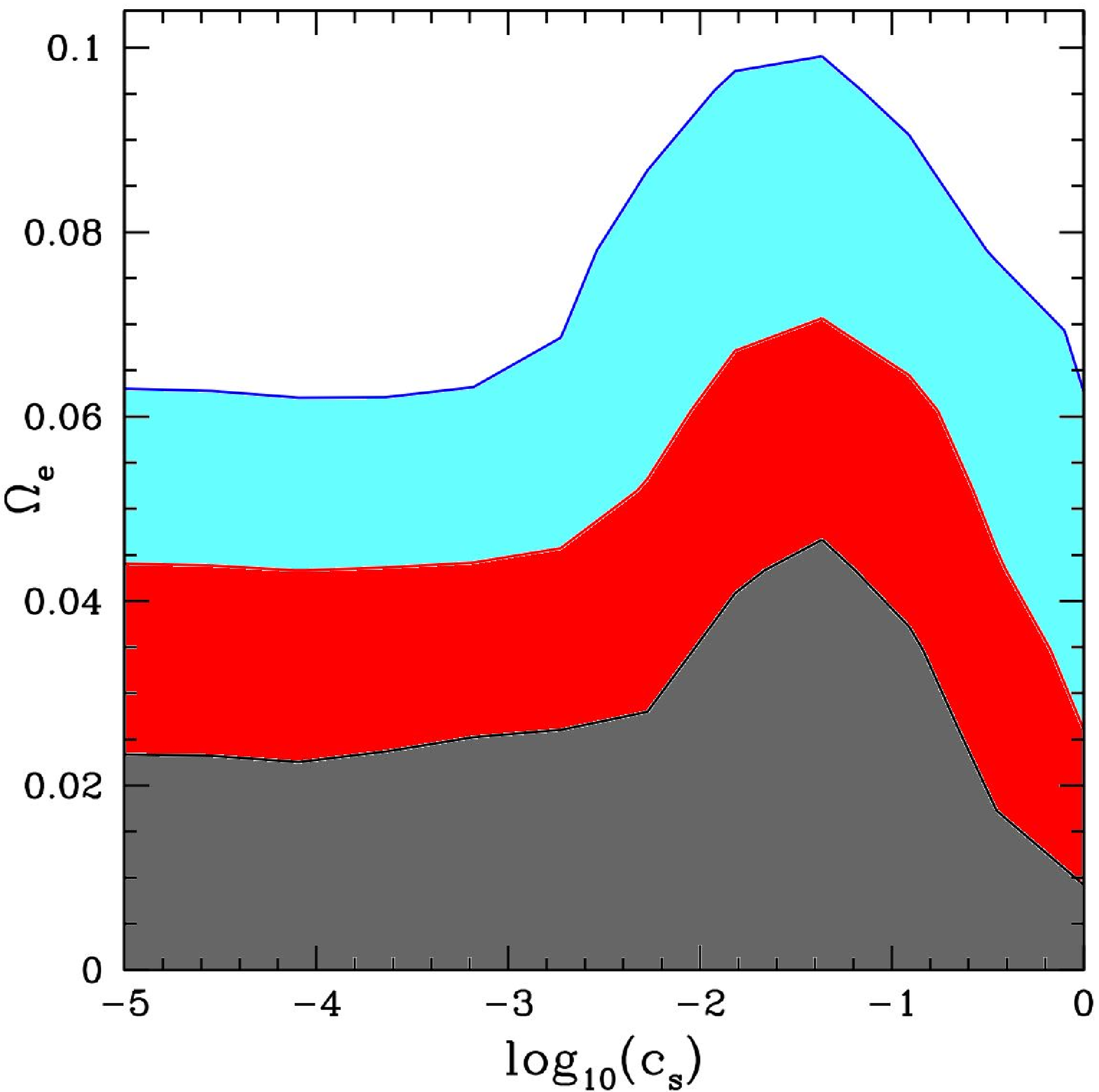,width=8.5cm}
  \caption{68.3, 95.4 and 99.7\% confidence level contours in the early 
dark energy model with constant sound speed $c_s$ and early dark energy 
density fraction $\Omega_e$.  The constraints are 
based on current data including CMB, supernovae, LRG power spectrum and 
crosscorrelation of CMB with matter tracers.  From \citet{sound}.} 
  \label{fig:soundw}
  \end{center}
\end{figure}

Theories along the lines of DBI or barotropic dark energy seem promising 
guideposts to a natural physical origin for acceleration, at least within 
the ``new component'' approach to dark energy.  Interestingly, both of 
them also make predictions for the microphysics of the dark energy distinct 
from simple scalar fields.  Minimally coupled, canonical (standard kinetic 
term) scalar fields have a sound speed of field perturbations equal to 
the speed of light, and hence do not cluster except on near horizon scales. 
Both the DBI and barotropic theories have sound speeds that instead 
approach zero (and hence could cluster) for at least part of their 
dynamics.  We explore this further in the next section, but 
Figure~\ref{fig:soundw} shows limits due to current data on the sound 
speed $c_s$ for a barotropic-type model (DBI models have even weaker 
constraints on $c_s$).

\section{The Frontiers of Nothing \label{sec:front}} 

From the previous section it may seem like our knowledge of nothingness 
is close to nothing.  But the past dozen years of experimental work and 
theoretical investigation have ruled out large classes of models, albeit 
perhaps the simplest ones and of course the ones with the most obvious 
experimental signatures.  The increasing difficulty has led some to 
pessimism, but the advancing network of diverse observational probes to 
be carried out over the next dozen years can be a source of hope. 

Recall the story of Auguste Comte, who in 1835 declared that ``we shall 
never be able to know the composition of stars''.  It was only 14 years 
later that it was discovered that the spectrum of electromagnetic 
radiation encodes the composition of material.  Perhaps within the next 14 
years there will be an analogous breakthrough of theoretical and 
experimental techniques for dark energy.  Such progress in scientific 
discovery has become gratifyingly habitual, as witness Richard Feynman's 
quote: 

\begin{quote}
Yesterday's sensation is today's calibration and 
tomorrow's background. 
\end{quote} 

\noindent Just as perturbations in the cosmic microwave background (CMB) 
radiation 
were undetected (and beginning to be despaired of) in the 1980's, 
discovered in the 1990's, and are today sometimes regarded as ``background 
noise'' relative to the signatures of galaxy cluster physics, so may 
the homogeneous background of dark energy and the value of $w(a)$ be 
treated in the future. 

What lies beyond $w$?  Even for the expansion history $a(t)$, i.e.\ the 
homogeneous dynamics of expansion, there is the question of whether 
dark energy makes a contribution at high redshift, whether in an 
accelerating form or not.  This is called early dark energy and current 
constraints are at the few percent level \citep{drw} -- by contrast the 
cosmological constant would 
contribute a fractional density of $10^{-9}$ at the CMB last scattering 
surface at $z\approx1090$.  Within a few years, CMB data from the Planck 
satellite should tighten the constraints by a factor 10.  

There is the 
issue of whether dark energy interacts with any other component other 
than through gravity.  This could become apparent through a situation 
such as cosmological neutrino mass bounds being at variance with laboratory 
measurements (if dark energy interacts with neutrinos, 
e.g.\ \citet{amennu,wettnu}), 
or through features in the matter density perturbation power spectrum (if 
dark energy interacts with dark matter, e.g.\ \citet{instab}). 

Does dark energy cluster?  This could come about either through a low 
sound speed (although it also requires that $w$ deviate appreciably from 
$-1$) or a coupling to other components.  Observationally this can be 
probed through detailed measurements of matter clustering on various 
length scales, using the next generation of galaxy surveys. 

Perhaps the most intriguing possibility is new laws of physics: in the 
``Neptune vs.\ post-Newton'' alternative to end up with extensions to 
the laws of gravitation beyond Einstein's general relativity rather than 
a new quantum scalar field.  It is not easy to find viable theories of 
gravity that accord with observations, and most of the ones that do exist 
are driven toward similarity with general relativity (GR).  Again, we seek a 
model independent approach that might identify some key features that a 
fundamental extended theory would need. 

The simplest generalization is to take a phenomenological approach of 
asking what feature of the observations could be shifted by a non-GR 
theory.  As mentioned in \S\ref{sec:dyn}, 
any modification of the expansion history is identical to an 
effective $w(a)$, so we must look further for an observational distinction. 
General relativity predicts a definite relation between the expansion 
history of the homogeneous universe and the growth history of energy 
density perturbations.  Other theories of gravity may deviate from this 
relation so we can define a gravitational growth index $\gamma$ that 
accounts for effects on growth beyond the expansion influence, seeing if 
it is consistent with the GR prediction. 

Parametrization of the growth of linear matter density perturbations 
$\delta\rho$ can be written as 
\beq 
g(a)=e^{\int_0^a (da'/a')\,[\om(a')^\gamma-1]} \,, 
\eeq 
where $g(a)=(\delta\rho/\rho)/a$.  This separates out the expansion 
history (which enters $\om(a)$) effects on growth from any extra 
gravitational influences (entering $\gamma$).  The gravitational 
growth index $\gamma$ is substantially independent of other cosmological 
parameters and can be determined accurately.  This form of representing 
deviations through $\gamma$, a single constant, reproduces the growth 
behavior to within 0.1\% accuracy for a wide variety of models 
\citep{groexp,lincahn}.  Note that other changes 
to the gravitational driving of growth besides the theory of gravity, 
such as other clustering components or couplings, can also cause $\gamma$ 
to deviate from its standard general relativity value of $0.55$. 

Moreover, gravitational modifications do more than affect growth: they 
alter the light deflection law in lensing and the relation between the 
matter density and velocity fields.  This can introduce both time and 
scale dependent terms.  In particular, the two potentials, appearing in 
the time-–time and space-–space terms of the metric, may no longer be 
equal as they are in general relativity, and the Poisson-type equations 
connecting them to the matter density and velocity fields could change. 
Among other approaches (e.g.\ \citet{gdm,husaw}), one can define new functions 
to account for these differences as \citep{danlin} 
\beqa 
-k^2(\phi+\psi)&=& 8\pi G_N a^2\bar\rho_m \Delta_m \times 
\mathcal{G} \label{eq:gdef}\\ 
-k^2\psi&=& 4\pi G_N a^2\bar\rho_m \Delta_m \times \mathcal{V} \,, 
\eeqa 
where $\phi$ and $\psi$ are the metric potentials, $\bar\rho_m \Delta_m$ the 
gauge invariant matter density perturbations, and $G_N$ is Newton's constant. 
In general relativity, the time and scale dependent functions $\mathcal{G}$ 
and $\mathcal{V}$ are identically unity. 

\begin{figure}[!t]
\begin{center} 
\psfig{file=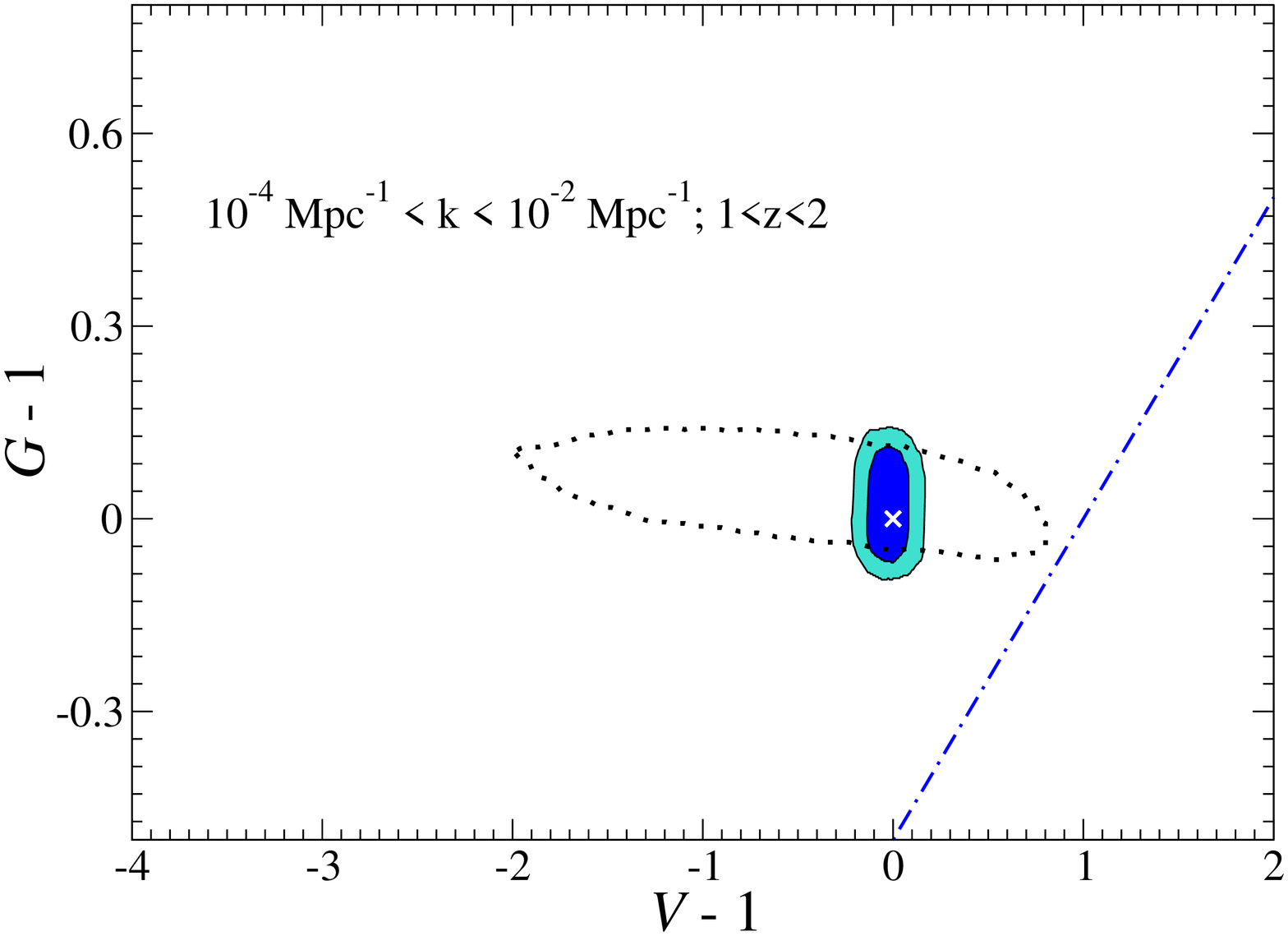,angle=-90,width=0.49\textwidth} 
\psfig{file=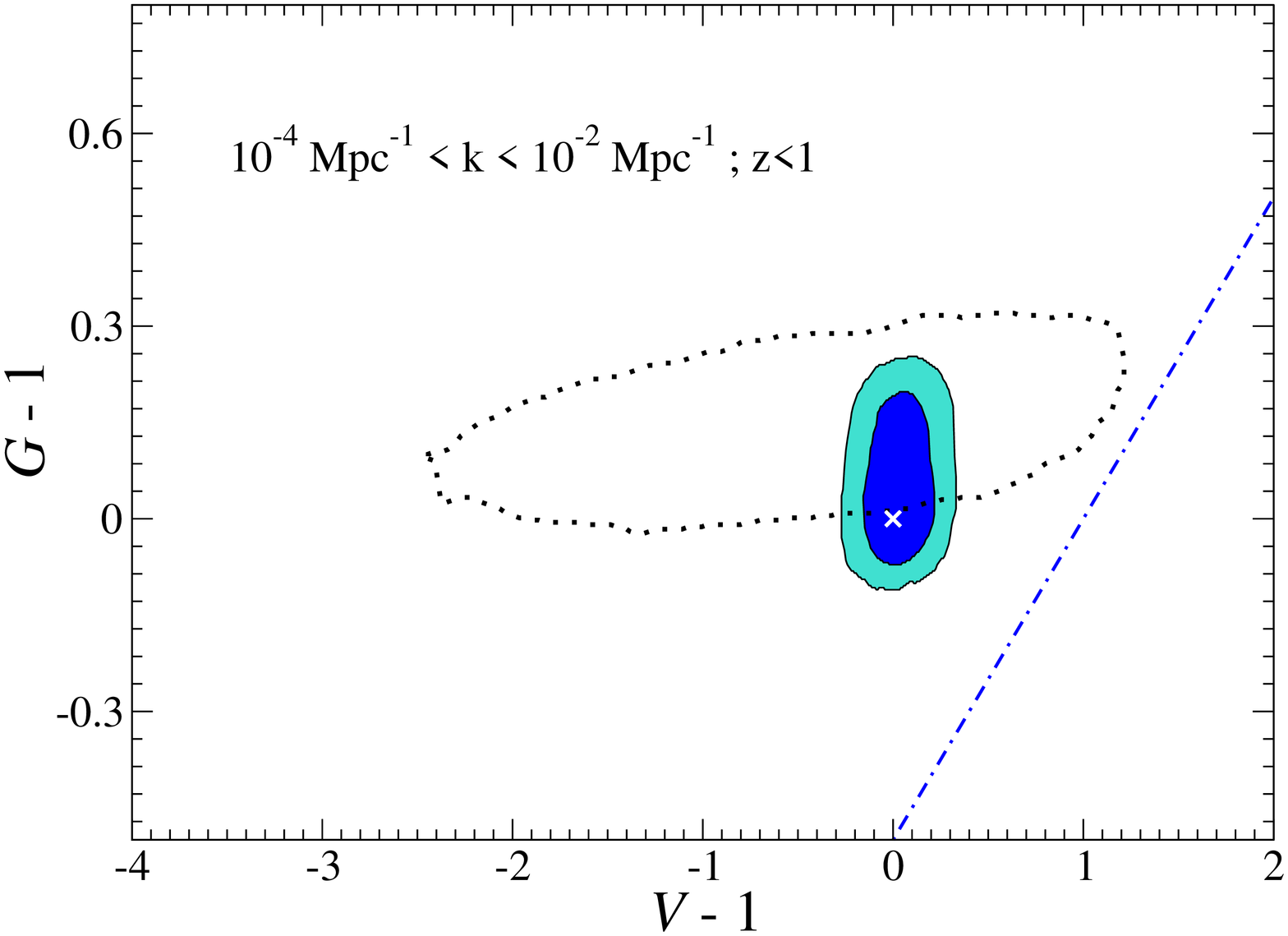,angle=-90,width=0.49\textwidth}\\ 
\psfig{file=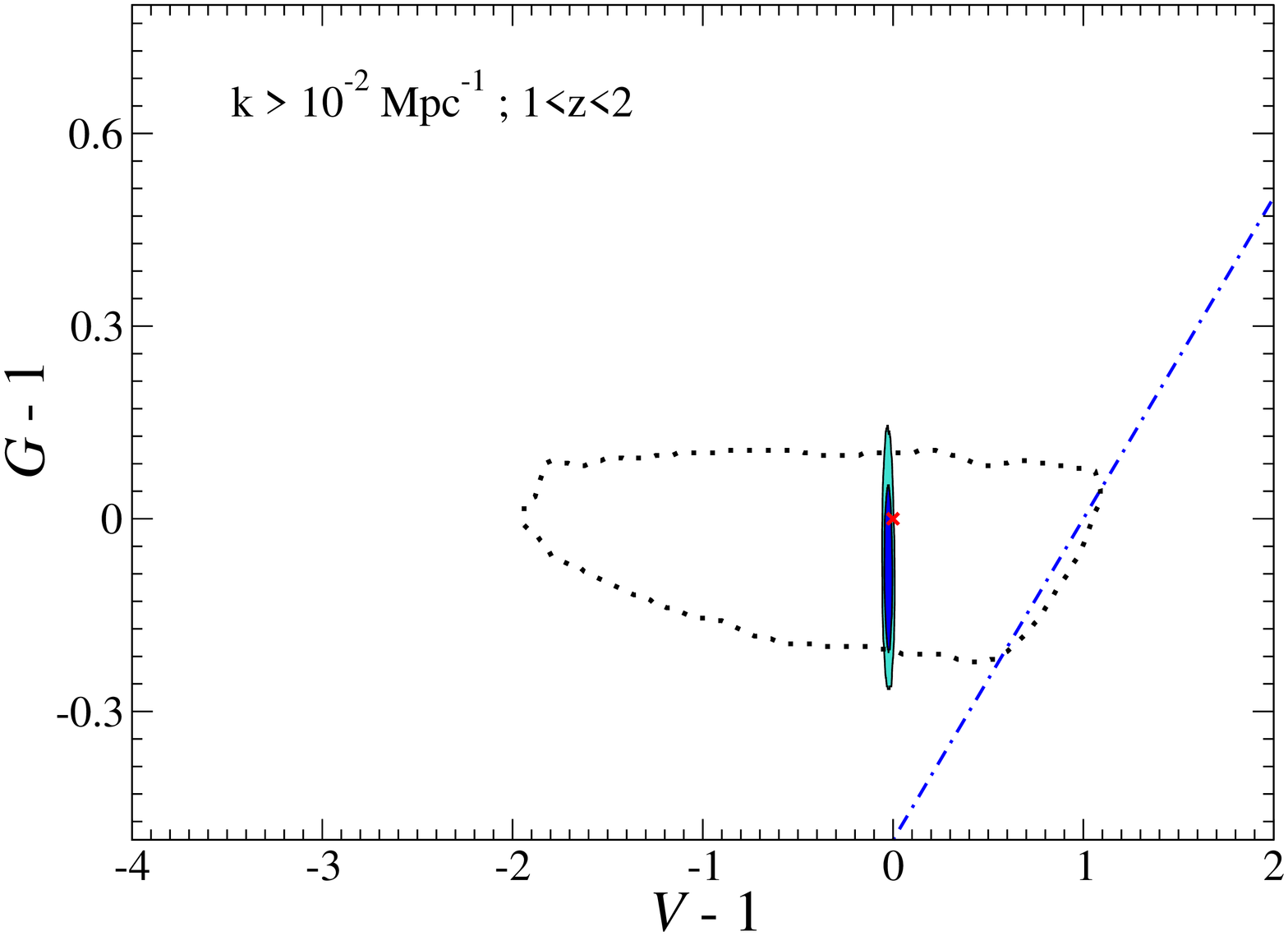,angle=-90,width=0.49\textwidth} 
\psfig{file=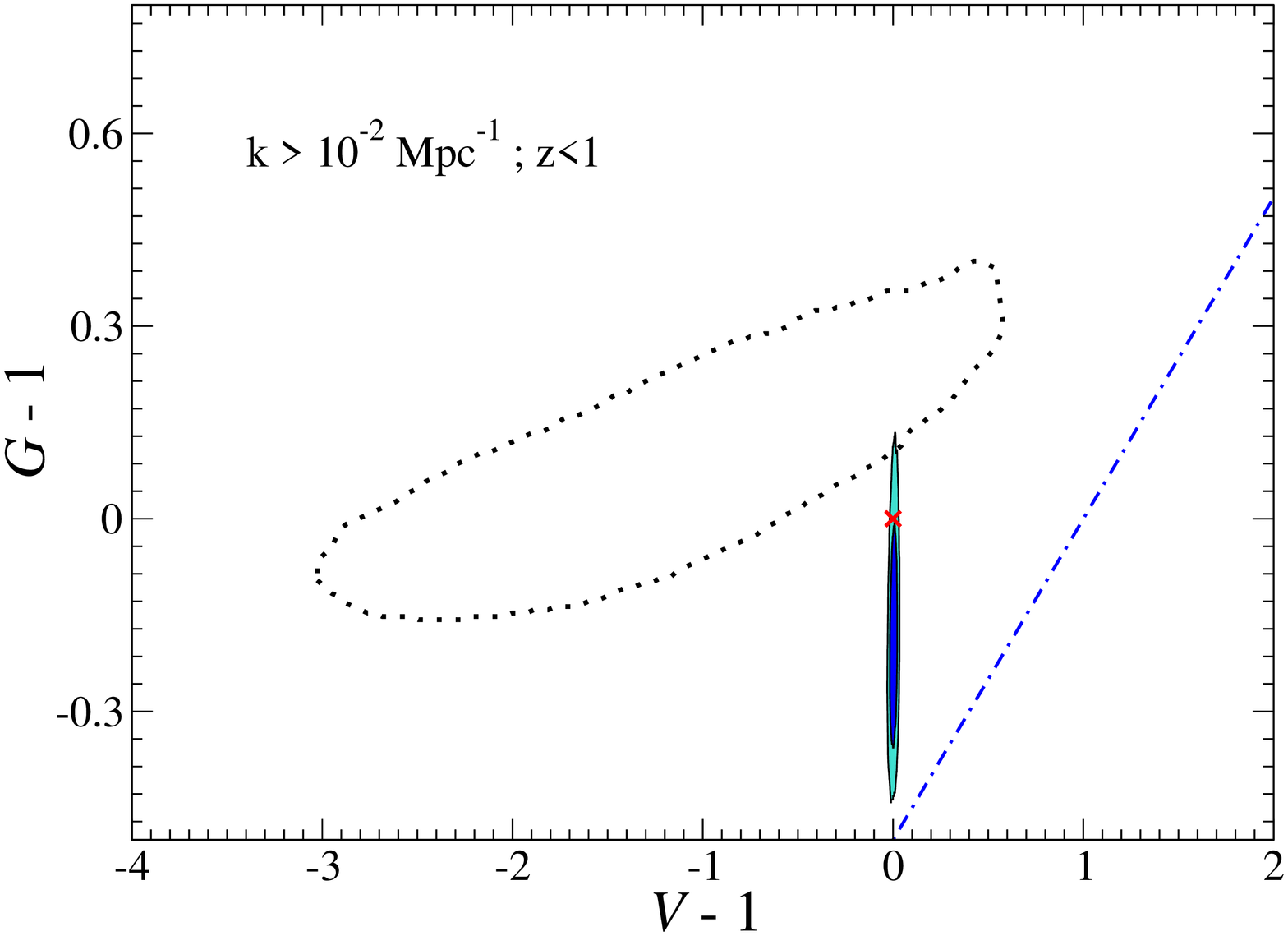,angle=-90,width=0.49\textwidth} 
\caption{
Filled contours show 68\% and 95\% cl constraints on $\vscr-1$ and 
$\gscr-1$ for the two redshift and two wavenumber bins using mock future 
BigBOSS, Planck, and JDEM supernova data.  The dotted contours recreate 
the 95\% cl contours from Figs.~8 of \citet{danlin} using current data 
(note the offset from (0,0) may be from systematics within the CFHTLS 
weak lensing data) to show the expected improvement in constraints.  
The x's denote the fiducial GR values.  Adapted from \citet{danlin}.} 
\label{fig:testgrav} 
  \end{center}
\end{figure}

Within a given theory of gravitation, the deviations $\gscr$ and $\vscr$ 
will be specified, but if we are searching for general deviations from 
Einstein gravity then we should take model independent forms for these 
functions.  Allowing their values to float in bins in redshift and in 
scale (wavenumber) gives considerable freedom and does not prejudice the 
search for concordance or contradiction with general relativity. 
Figure~\ref{fig:testgrav} shows both the current constraints and those 
expected from the next generation galaxy redshift surveys.

Considerable current data exists to constrain gravity and cosmology, 
including the cosmic microwave background (CMB), supernova distances, 
weak gravitational lensing, galaxy clustering statistics, and 
crosscorrelation between the CMB photon 
and galaxy number density fields.  Nevertheless, although this now 
constrains the sum of the potentials, and hence $\gscr$ (see 
Eq.~\ref{eq:gdef}), fairly well, the growth of structure, in terms of 
$\vscr$, is still poorly known.  This should change with the next 
generation of large volume, three dimensional galaxy mapping surveys. 
We see that the 8 ``beyond GR'' gravity parameters ($\gscr$ and $\vscr$ 
each in two redshift and two wavenumber bins) could be determined to 
within $\sim10\%$ or better. 

Testing gravity on cosmic scales is an area of intense interest at the 
moment; previous work using current data \citep{daniel10,bean10,zhao10,uros10} 
finds consistency with GR, although again deviations are certainly 
allowed.  Better data from growth probes could play a key role in tightening 
constraints or uncovering new physics.   A particularly exciting 
prospect is comparing the density, velocity, and potential field 
information through combining imaging and spectroscopic surveys 
\citep{jainzhang,uros10,jainkhoury}.  Extending our probes and 
understanding into the nonlinear density structure regime is another 
area of active exploration \citep{oyaizu,schmidt}.

\section{Conclusions} 

Beyond the atoms and photons that make up our familiar world, and 
all the particles of the Standard Model of particle physics, the 
nature of the vacuum and spacetime is a mystery that has come to 
the forefront of physics.  Gravity, the most familiar and omnipresent 
of all the forces, is not behaving as we expected.  More than 70\% 
of the energy density in the universe is made of nothing -- nothing 
we have experienced before.  Conditions are ripe for a true adventure 
in cosmology. 

While current data are consistent with a cosmological constant as 
a source for dark energy, a cornucopia of other physical origins 
are in agreement as well.   We do not yet know whether the dark energy 
is uniform, is dynamic, disappears at early times, is of a quantum origin 
or a gravitational origin.  All are valid possibilities, and carry 
profound implications for the frontiers of physics and the fate of 
the universe. 

A key question is whether we are dealing with a new physical ingredient 
or new physical laws -- or both.  For example, the dark energy may 
interact with neutrinos through a novel interaction; is dark energy 
really a completely separate sector of physics or are there new 
forces and symmetries as intricate as in known particle physics? 
We are very much at the beginning of our explorations 
of the frontier physics of dark energy and cosmic acceleration.

The exciting goal of future observations is to explore this 
wonderland of physics.  We have few robust models but some general 
concepts, and some excellent model independent parametrizations.  
For the dynamical aspects of cosmic expansion, 
next generation measurements of the equation of state and its time 
variation, $w$ and $w'$, in the calibrated form of $w_0$ and $w_a$ 
describe the experimental reach to the subpercent level of 
observational accuracy. 
Comparison of tests of growth and expansion could give key clues to 
the underlying physics, as can contrasting the density, velocity, 
and gravitational potential fields of large scale structure.  These 
should be enabled by a diverse network of future observations, 
delineating the physical properties of dark energy and testing general 
relativity.  At the same time, these measurements deliver information 
of great value to many 
other astrophysical explorations as we map the structure, motion, 
and growth in our universe. 

Settling the frontier will require challenging efforts by both observers 
and theorists.  One must not only measure the expansion history $w(a)$, 
growth history $\gamma$, gravity $\gscr$ and $\vscr$, couplings, early 
dark energy etc. -- but also understand them.  Even if we fail to 
detect deviations from a cosmological constant, we cannot say the 
revolutionary physics of dark energy is known until we explain it.  
As two British Astronomers Royal said in the 19th century: 

\begin{quote} 
``I should not have believed it if I had not seen it!'' -- Sir G.B.\ Airy
\end{quote} 
and the reply 
\begin{quote} 
``How different we are! My eyes have too often deceived me.  I believe 
it because I have proved it.'' -- Sir W.R.\ Hamilton
\end{quote}

\section*{Acknowledgments} 

I thank Dragan Huterer for useful comments and 
the Centro de Ciencias Pedro Pascual in Benasque, Spain for 
hospitality during the writing of part of this chapter.  
This work has been supported in part by the Director, Office of Science, 
Office of High Energy Physics, of the U.S.\ Department of Energy under 
Contract No.\ DE-AC02-05CH11231, and World Class University grant 
R32-2009-000-10130-0 through the National Research Foundation, Ministry 
of Education, Science and Technology of Korea.

\end{document}